\documentstyle[aps,epsf]{revtex}  % DO NOT DELETE THIS LINE 
\newcommand{\shit}[2]{\epsfxsize=#1 \epsfbox[10 30 560 590]{./#2}}
\begin{document}        %  DO NOT DELETE OR CHANGE THIS LINE
%%
%% some of my defs.
%%
\newcommand{\etal}{{\it et al.} }
\newcommand{\ie}{{\it ie.} }
\newcommand{\eg}{{\it eg.} }
\newcommand{\vs}{{\it vs.} }
\newcommand{\plus}{\makebox[15pt][c]{$+$}}
\newcommand{\minus}{\makebox[15pt][c]{$-$}}
\newcommand{\er}[2]
{\hskip-0.5em\raisebox{0.08em}{\scriptsize{$\;\begin{array}{@{}l@{}}
\plus\makebox[0.25em][r]{#1\hfill} \\[-0.12em]
\minus\makebox[0.25em][r]{#2\hfill} 
\end{array}$}}}
\newcommand{\err}[2]
{\hskip-0.5em\raisebox{0.08em}{\scriptsize{$\;\begin{array}{@{}l@{}}
\plus\makebox[0.50em][r]{#1\hfill} \\[-0.24em]
\minus\makebox[0.50em][r]{#2\hfill} 
\end{array}$}}}
\newcommand{\errr}[2]
{\hskip-0.5em\raisebox{0.08em}{\scriptsize{$\;\begin{array}{@{}l@{}}
\plus\makebox[0.80em][r]{#1\hfill} \\[-0.48em]
\minus\makebox[0.80em][r]{#2\hfill} 
\end{array}$}}}
\newcommand{\errrrr}[2]
{\hskip-0.5em\raisebox{0.08em}{\scriptsize{$\;\begin{array}{@{}l@{}}
\plus\makebox[0.80em][r]{#1\hfill} \\[-0.72em]
\minus\makebox[0.80em][r]{#2\hfill} 
\end{array}$}}}
\baselineskip 14pt
\title{Progress in leptonic and semileptonic decays in lattice QCD}
\author{Sin\'ead M.~Ryan}
\address{Fermi National Accelerator Laboratory,\\P.O. Box 500, Batavia, IL 60510\footnote{Fermilab is operated by the Universities Research Association under contract with the US Department of Energy}}
% \author{}   % Use this and the next line only if there is a second
% \address{Another University, etc.}  % address. (Remove the left % marks)
%
\maketitle              % Creates the title area, Do Not Remove
\begin{abstract}        % Do Not Delete this line
The status of lattice calculations of heavy quark phenomenology is reviewed. 
Particular emphasis is placed on the understanding and control of the calculational uncertainties.
The ensuing implications for constraining the CKM matrix elements are discussed.  
\end{abstract}   	% Do Not Delete this line
\section{Introduction} 
The calculation of hadronic matrix elements plays a vital r\^ole in 
determinations of the CKM matrix elements and overconstraining the unitarity
triangle of the Standard Model.
In particular, the CKM matrix elements that are combined to make up the sides 
of the triangle can be determined from neutral meson mixing and exclusive leptonic and semileptonic decay processes in the following way
\begin{equation}
\left\{ \begin{array}{c}\mbox{experimental measurement of}\\ \mbox{decay rate}\; or\; \mbox{meson mixing}\end{array}\right\} = \left\{ \begin{array}{c}\mbox{known} \\ \mbox{factors}\end{array}\right\}\left\{ \begin{array}{c}\mbox{nonperturbative}\\ \mbox{form factor}\; or\; \mbox{decay constant}\end{array}\right\}|V_{CKM}|^2 , \label{decay_cartoon}
\end{equation}
$|V_{CKM}|$ can be determined by combining experimental measurements of exclusive decay rates with theoretical calculations of the nonperturbative contributions.
These calculations can be done with a number of methods including lattice QCD. 
Because the lattice formalism of QCD is model independent and systematically improvable it should offer some of the most reliable results.            
Currently, the uncertainty in many of the theoretical and experimental inputs is large enough that unitarity is not really tested. 
Table~\ref{CKM_tab} shows the decay modes to be discussed here and the associated CKM matrix elements. Indeed $V_{td}$ and $V_{ub}$ are amongst the least well determined components with much of the error coming from the nonperturbative contributions~\cite{PDG}.
However, the next generation of B-physics experiments will reduce the experimental uncertainty considerably and to take full advantage of this the theoretical uncertainties must also be reduced.

The form factors and decay constants of Equation~\ref{decay_cartoon} are extracted from hadronic matrix elements, which can be calculated directly in lattice QCD. 
Lattice calculations of heavy quark quantities began in the late '80s and it was quickly realised that they could provide important inputs to phenomenologically interesting parameters of the Standard Model. 
Over the years the field has matured enormously as a result of a deep understanding of the particular systematic errors involved and a concerted effort to control or remove them, as described in Section II. 
The largest remaining uncertainty in the numerical results is the quenched approximation. As this is removed in the next several years the lattice results will be the results of QCD and not a model.
The aim of this paper is to give a progress report for heavy quark physics from lattice QCD emphasising the new results where attention has been paid to the particular systematic uncertainties involved. The state-of-the-art leptonic decay constants are described in some detail and 
a brief report on some new ideas for semileptonic physics is included. A more extensive review can be found in Ref.~\cite{draper}.
\begin{table}[tb]
\caption{The CKM matrix elements and nonperturbative (lattice) calculations which are discussed in the text.}
\begin{tabular}{cccc} 
CKM Matrix element & Process & Lattice Calculation \\
\tableline 
\tableline	
$|V_{td}|$ & $B^0_d-\bar{B}_d^0$ mixing & $f_B\sqrt{B_d}$\\
$|V_{ts}|$ & $B^0_s-\bar{B}_s^0$ mixing & $f_{B_s}\sqrt{B_s}$\\
$|V_{ub}|$ & $B\rightarrow\pi l\nu$ & $f_{\pm }^{B\rightarrow\pi}(q^2)$  \\
$|V_{cb}|$ & $B\rightarrow Dl\nu$ & $h_{\pm}(\omega )$\\ 
	   & $B\rightarrow D^\ast l\nu$ & $h_{A_1}(\omega )$  \\
\end{tabular}
\label{CKM_tab}
\end{table}
\section{Progress in heavy quark physics}
\begin{figure}[ht]	% in second brace, h=here, t=top, b=bottom	
\centerline{\epsfxsize=4in\epsfysize=2.6in\epsfbox{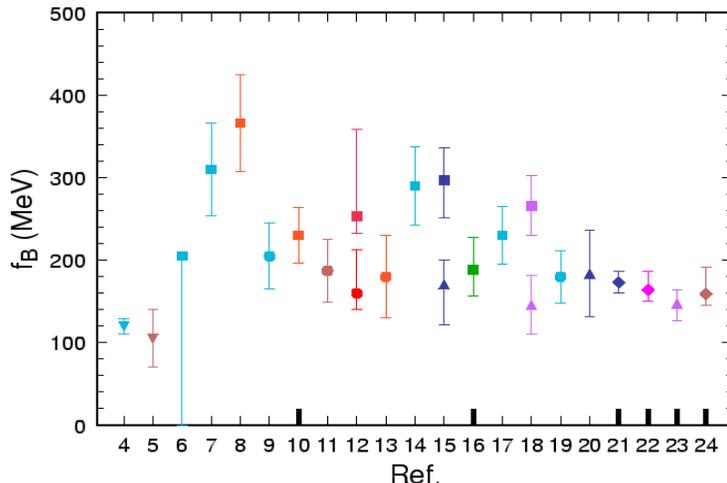} }
\vskip -.2 cm
\caption[]{   
\label{askfB}
\small $f_B$ vs time-ordered journal reference, as in Ref.~\cite{askHQ_proc}. Symbols represent the different methods for heavy quarks used: $\bigtriangledown$ is extrapolation from $m_{ch}\rightarrow m_b$; $\Box$ is $m_Q\rightarrow\infty$; $\bigcirc$ is interpolation $[m_{ch},\infty]$; $\bigtriangleup$ is NRQCD and 
$\Diamond$ is the Fermilab approach.}
\end{figure}

The evolution of lattice calculations using heavy quarks is shown very nicely in a plot made by Andreas Kronfeld for the Heavy Quark '98 Workshop~\cite{askHQ_proc}. Figure~\ref{askfB} shows the lattice determinations of $f_B$ from the earliest calculations to the most recently published ones, 
demonstrating how calculations of $f_B$ have matured over the years. Advances in computing power and our understanding of heavy quark physics mean the central value and error bars have settled down remarkably. The last four points are the most recent values of $f_B$ from the JLQCD~\cite{JLQCDfB}, Fermilab~\cite{fB_paper}, NRQCD~\cite{GLOKfB} and MILC~\cite{MILCfB} collaborations respectively. 
I also note the UKQCD~\cite{UKQCDfBdgr,UKQCDfBlin} and APE~\cite{APEfB} collaborations have new determinations of $f_B$ which are not included in this plot but which will be included in the text.
The remainder of this section details the dominant uncertainties in a lattice calculation of heavy quark matrix elements and how they are treated in different approaches.
\subsection{${\cal O}(am_Q)$ discretistion errors}
Present day computing resources mean lattice calculations are generally done in a box of size $\leq 3$ fm and with a lattice spacing $a\approx 0.07 - 1.2$ fm. Simulating realistically heavy quark masses at these lattice spacings must be approached carefully due to discretisation errors proportional to the heavy quark mass. A brute force reduction of $a$ is extremely costly, requiring orders of magnitude more computing power so improving lattice actions is a necessary step.
One approach is to use an improved light-quark action, with the improvement coefficients of the action and operators calculated nonperturbatively. 
Simulations are done with quark masses around that of charm. Heavy quark effective theory is then used to guide an extrapolation from this region to that of the bottom quark. In some cases results of a calculation in the static approximation can be used - allowing an interpolation in quark mass instead of an extrapolation. Both UKQCD and APE collaborations use this method. However, this extrapolation to a region where ${\cal O}(am)$ errors are not under control introduces a systematic error which is hard to quantify.

A different approach is to consider the B system as nonrelativistic: for $b$ quarks $(v/c)^2\approx (0.3{\rm GeV}/5.0{\rm GeV})^2$. Relativistic momenta are excluded by
introducing a finite cut-off such that $p\approx m_Qv\ll m_Q$. Then, $|p|/m_Q\ll 1$ and the QCD Lagrangian can be expanded in powers of $1/m_Q$~\cite{NRQCD}. This has been extremely successful for B physics, where the quark mass is large, so the expansion is convergent. Results from the GLOK group using NRQCD are shown here. The theory is nonrenormalisable which means a formal continuum limit ($a\rightarrow 0$) does not exist so results must be obtained at finite lattice spacing. Lattice artefacts are removed by including higher orders in $1/m_Q$ and $a$.

The final approach is that of the Fermilab group~\cite{KKM}. It is a 
re-interpretation of the relativistic action which identifies and correctly 
renormalises nonrelativistic operators present in the {\it so called} light-quark action. Discretisation errors are then ${\cal O}(a\Lambda_{QCD})$ and not ${\cal O}(am_Q)$. Also the existence of a formal continuum limit means a continuum extrapolation is possible. 
The result is an action valid for arbitrary $am$ and for 
which the systematic errors can be understood and controlled thereby allowing simulations at the $b$ quark mass. 
This approach is used by Fermilab, JLQCD and MILC.
\subsection{Lattice spacing dependence}
If the large mass-dependent errors are under control the remaining lattice spacing dependence can be quantified by repeating the matrix element calculation at three or more lattice spacings and extrapolating the result to the continuum ($a=0$) limit. 
This has been done by the collaborations using the Fermilab approach in their calculations of heavy-light decay constants. 
So far it has only been done by one group for the semileptonic case~\cite{my_lat98}. In general, only a mild dependence on lattice spacing is observed and 
reliable extrapolations can be performed. 
\subsection{Operator matching}
Matrix elements determined on the lattice must be related to their continuum counterpart as in
\begin{equation}\nonumber
\langle P^\prime (\vec{k})|{\cal O}_\mu|P(\vec{p})\rangle_{continuum} = 
{\cal Z}_\mu (\mu ,g^2)\langle P^\prime (\vec{k})|{\cal O}_\mu|P(\vec{p})\rangle_{lattice} .
\end{equation}
The renormalisation factor, ${\cal Z}_\mu$ is usually calculated in perturbation theory and has some associated uncertainty.
UKQCD and APE use a nonperturbative determination of both the ${\cal Z}_\mu$ and the improvement coefficients in their actions, as described in~\cite{alpha_collab}, which reduces considerably the matching uncertainty.
\subsection{Quenching}
Most lattice calculations are done in the quenched approximation (omitting light quark loops) as a computational expedient. The uncertainty introduced by this approximation is difficult to quantify. There are now however, some partially unquenched calculations of heavy quark quantities and these have been used by many collaborations to estimate the effect of quenching on their results.
%%%
%
%
\section{Leptonic Decay Constants}
The calculation of heavy-light leptonic decay constants is an important application of lattice techniques for two reasons. Firstly, the decay constants themselves are important phenomenological parameters. In particular, the $B$ meson decay constant, which has not been measured experimentally, is an input in determinations of the CKM matrix element $V_{td}$ through $B^0_d-\bar{B}^0_d$ mixing,
\begin{equation}
\Delta m_d(B^0_d-\bar{B}^0_d) = \left [\frac{G_F^2m_W^2}{6\pi}\eta_{B_d}{\cal S}\left (\frac{m_t}{m_W}\right )\right ]f_{B_d}^2\hat{B}_{B_d}|V_{td}V_{tb}^\ast |^2 .
\end{equation}
$\Delta m_d$ has been measured experimentally to $\approx 4\%$~\cite{PDG} and 
the factor in square brackets is also well determined.  
Interestingly $V_{td}$ is known to only $\approx 25\%$ accuracy and the bulk of that uncertainty comes from the nonperturbative input $f_B^2\hat{B}_{B_d}$. The bag parameter, $\hat{B}_{B_d}$ is also calculated on the lattice but will not be discussed here, some recent reviews are in Refs. ~\cite{draper} and ~\cite{askHQ_proc}.  

Secondly, the calculation of leptonic decay constants is an important test-ground for lattice heavy quark techniques. Control of systematic errors here inspires confidence in the more complicated determinations of form factors and partial widths in the semileptonic case. 

\begin{figure}[ht]	% in second brace, h=here, t=top, b=bottom	
\centerline{\epsfxsize=4in\epsfysize=2.6in\epsfbox{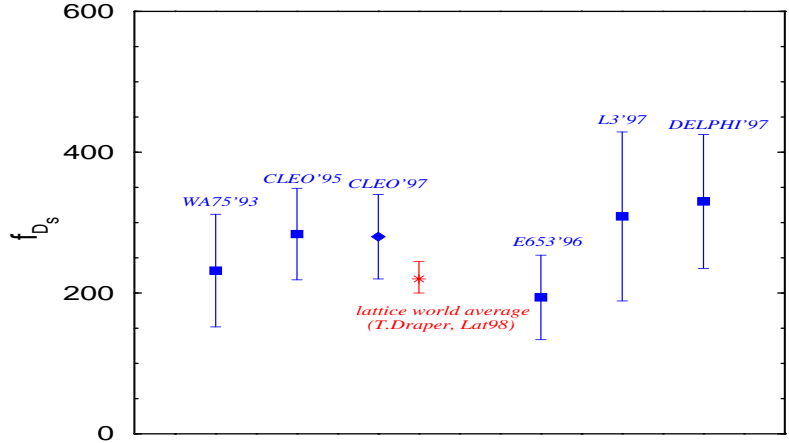}}
\vskip -.2 cm
\caption[]{
\label{fDswrl}
\small Comparing the world-average lattice result for $f_{D_s}$ with the experimental data~\cite{WA75,CLEO,E653,L3,DELPHI}.}
\end{figure}

Naturally a calculation of the $B$ meson decay constant can also include the decay constants of the $B_s$ meson and the $D$ system. A number of experiments have measured $f_{D_s}$ so the lattice result can be compared with the experimental numbers for at least one heavy-light decay constant. The agreement, shown in Figure~\ref{fDswrl} should inspire confidence in the determination of $f_B$.
In recent years there have been a number of new calculations of $f_B$ using different approaches described in Section II, however a common feature is a comprehensive treatment of the systematic uncertainties involved. 
Table~\ref{fB_tab} lists results from the most recent calculations of the heavy-light decay constants and one can see that they are in rather good agreement now. World averages for all these decay constants can be found in the review by Draper~\cite{draper}. 

\begin{figure}[ht]	% in second brace, h=here, t=top, b=bottom	
\centerline{\epsfxsize=4in\epsfysize=2.6in\epsfbox{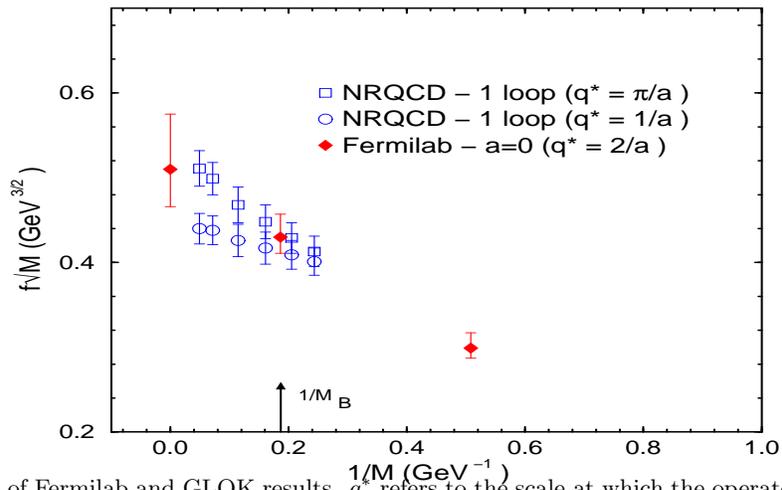} }
\vskip -.2 cm
\caption[]{
\label{compare_nrqcd}
\small A comparison of Fermilab and GLOK results. $q^\ast$ refers to the scale at which the operator matching is done and the Fermilab result is in the range considered by GLOK. Results from GLOK are at $\beta=6.0$, Fermilab results are in the continuum limit.}
\end{figure}

\begin{figure}[ht]	% in second brace, h=here, t=top, b=bottom	
\begin{center}
\setlength{\unitlength}{1cm}
\setlength{\fboxsep}{0cm}
\begin{picture}(14.5,8)
\put(0,2){\begin{picture}(7,7)\put(-0.9,-0.4)
{\shit{6.5cm}{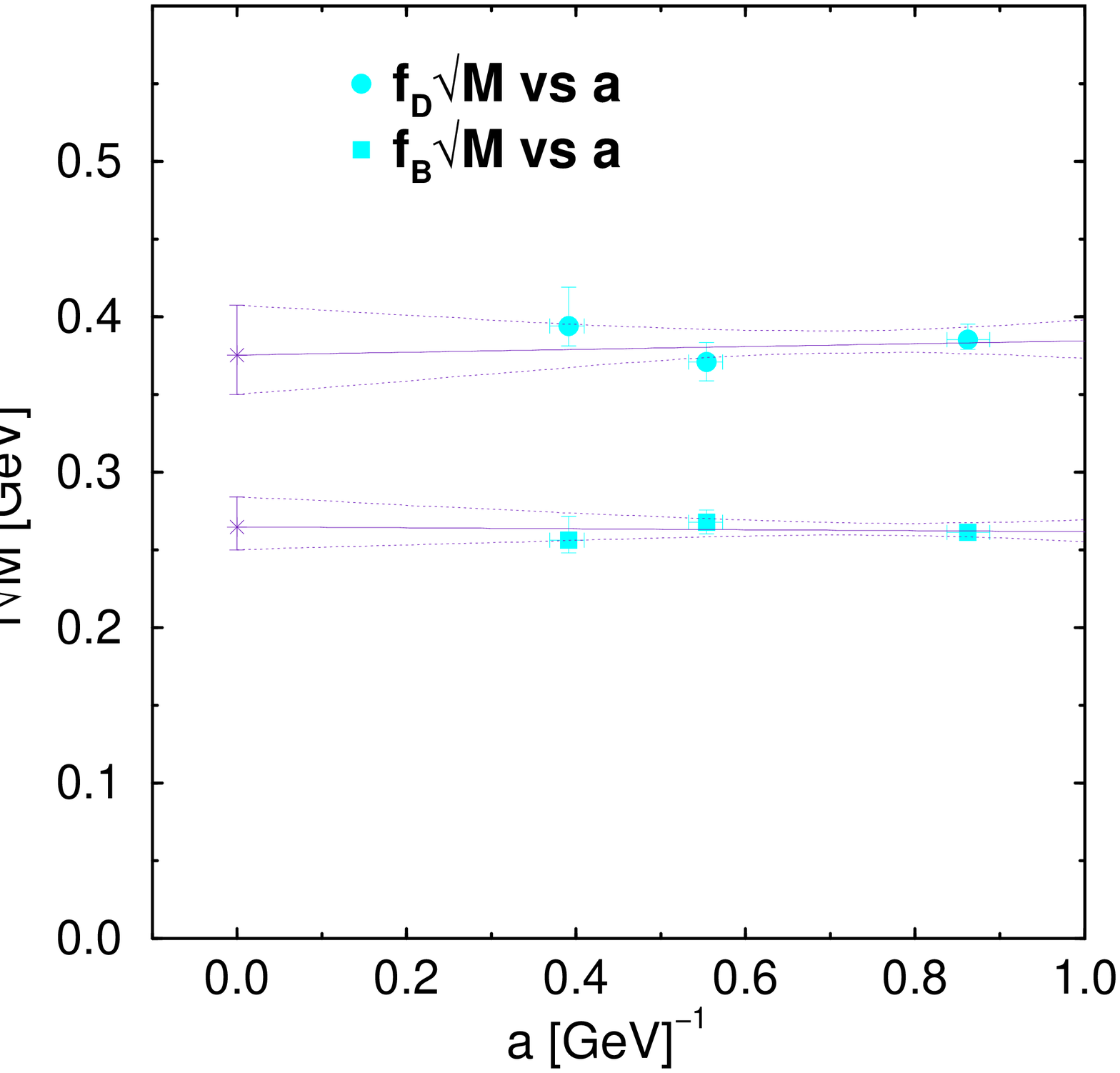}}\end{picture} }
\put(7.5,2){\begin{picture}(7,7)\put(-0.9,-0.4)
{\shit{6.5cm}{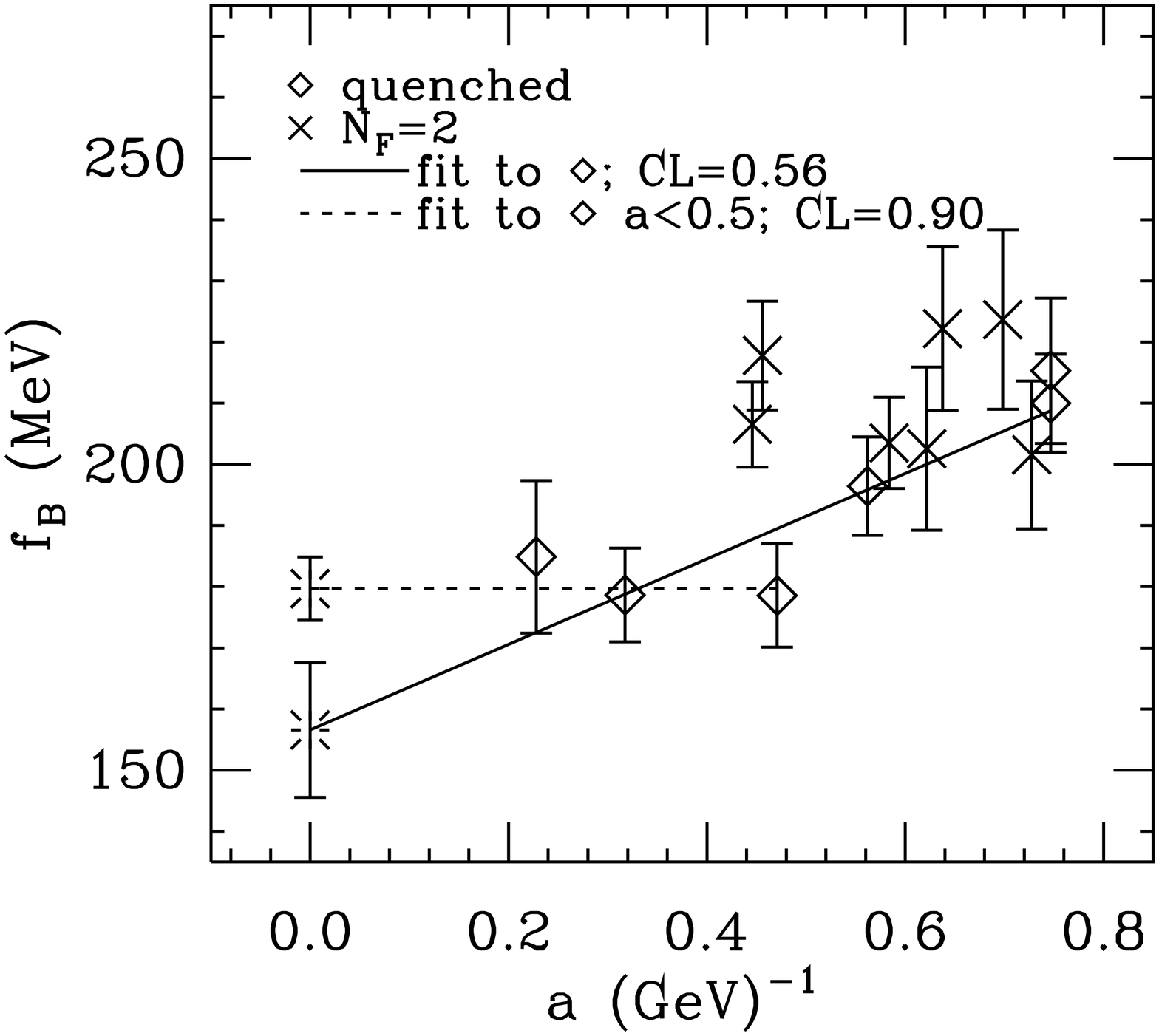}}\end{picture}}
\end{picture}
\end{center}
\vskip -.2 cm
\caption[]{
\label{fB_aeq0}
\small Continuum extrapolations: Fermilab extrapolates $f\sqrt{M}$ to $a=0$ (left plot) and MILC extrapolates $f_B$ (right plot). The Fermilab action has a milder dependence on $a$ as expected since the action is more highly improved.}
\end{figure}

Since both NRQCD and the Fermilab approach work well at the $B$ meson mass it is interesting to compare the results. Figure~\ref{compare_nrqcd} shows that for $f\sqrt{M}$ (the amplitude of the matrix element) at $1/m_B$ there is good agreement .
To illustrate the care taken in most calculations, the systematic error quoted by the Fermilab group was arrived at by considering the following:
\begin{tabbing}
\hspace{1cm}\={\it (i)} Excited state contamination is less than statistics ;\=\hspace{0.5cm} \={\it (ii)} Finite volume is less than statistics\=\hspace{1cm} \\
\>{\it (iii)} Lattice spacing dependence is less than statistics ; \>\>{\it (iv)} Heavy quark tuning is less than statistics\> \\
\>{\it (v)} Perturbation theory = 5\% ;\>\> \>
\end{tabbing}
So the systematic error in this calculation is almost entirely from the perturbation theory used to match the lattice and continuum operators, as described in Section II. 
The pertubative calculation used is a one-loop, mass-dependent
result from Aoki {\it et al}~\cite{aoki_pertTh} for which the leading error is taken to be ${\cal O}(\alpha_s^2)$. 
The groups who studied the dependence on lattice spacing found it to be gentle, examples of the continuum extrapolations are shown in Figure~\ref{fB_aeq0}.
The quenching error has been estimated from work by the MILC collaboration who 
produced the first results for an unquenched $f_B$ using a relativistic action. They found an increase in $f_B$ of $\approx 10\%$ at a finite lattice spacing. Thus a ($+16$MeV) estimate from Fermilab. Other collaborations have used this work to make similar estimates (some taking an increased central value and symmetric error bars). Preliminary results for an unquenched $f_B$ after extrapolation to the continuum limit suggest that the effect may be larger than 10\% and closer to 25\%~\cite{bernard_nf2}. This is supported by the work of the 
 NRQCD group, who have recently estimated this effect to be $+25\%$~\cite{sara_nf2}.

The ratios of decay constants are also calculated. In a ratio many of the systematic errors may cancel leading to a more precise result. This can be used to place bounds on $|V_{td}/V_{ts}|$ and (assuming three generations) $|V_{td}|$ by using
\begin{equation}
\frac{\Delta m_{B_s}}{\Delta m_{B_d}} = \frac{m_{B_s}}{m_{B_d}}\frac{\hat{B}_{B_s}f_{B_s}^2}{\hat{B}_{B_d}f_{B_d}^2}\frac{|V_{tb}^\ast\cdot V_{ts}|^2}{|V_{tb}^\ast\cdot V_{td}|^2} .
\end{equation}
 \begin{table}[t]
 \caption{Recent calculations of $f_B$ and $f_{B_s}/f_B$ from different collaborations. The two preliminary results from UKQCD were obtained in separate analyses.}
 \begin{tabular}{ccccccc} % In second brace, l = left, r = right,
 Group & $f_B$ &$a^{-1}$ &Statistical Error & Systematic Error& Quenching & $f_{Bs}/f_B$\\  
       & [MeV] & [GeV]        &   [MeV]    &   [MeV]   &   [MeV]    & \\
 \tableline 
 \tableline	
 FNAL~\cite{fB_paper} & 164 & $a\rightarrow 0$ &\err{14}{11} & $\pm 8$ &\err{16}{0}& 1.13\err{5}{4} \\ 
 MILC~\cite{MILCfB} & 159 & $a\rightarrow 0$&$\pm 11$ &\err{22}{9} & \err{21}{0}& 1.11$\pm 2$\err{4}{3}$\pm 3$\\
 JLQCD~\cite{JLQCDfB}& 173 & $a\rightarrow 0$&$\pm 4$ & $\pm 9$ & $\pm 9$ & \\
 GLOK~\cite{GLOKfB} & 147 &1.92 &$\pm 11$ & $\pm 11$&\err{8}{12}& 1.20$\pm 4$\err{4}{0}\\
 \tableline
 UKQCD~\cite{UKQCDfBdgr,UKQCDfBlin} (prelim.)& 176, 161 &2.64, 2.58 &\err{5}{4}, $\pm 16$ & & &\\
 APE~\cite{APEfB} & 179 &2.75 &$\pm 18$ &\err{26}{9}& & 1.16$\pm 4$\\
 \end{tabular}
 \label{fB_tab}
 \end{table}
%
%%
%%%%%%%%%%%%%%%%%%
\section{Semileptonic B meson Decays}
The lattice treatment of semileptonic decays of heavy-light mesons is not yet as mature as that of the leptonic decays which is clear from the absence of a complete estimate of systematic errors \`a la $f_B$. The good news is that this is underway by a number of groups and should be available soon. 
In particular, calculations using NRQCD and the Fermilab approach herald a new era of precision in lattice calculations of form factors because the mass-dependent uncertainties are decoupled from other effects as a result of working at the $b$ quark mass. 
Only B meson decays will be discussed here but $D\rightarrow\pi l\nu$ and $D\rightarrow Kl\nu$ are also being studied~\cite{draper,simone_dpf}. 
\subsection{$B\longrightarrow\pi l\nu$}
\begin{figure}[ht]	% in second brace, h=here, t=top, b=bottom	
\begin{center}
\setlength{\unitlength}{1cm}
\setlength{\fboxsep}{0cm}
\begin{picture}(14.5,8)
\put(0,2){\begin{picture}(7,7)\put(-0.9,-0.4)
{\shit{6.5cm}{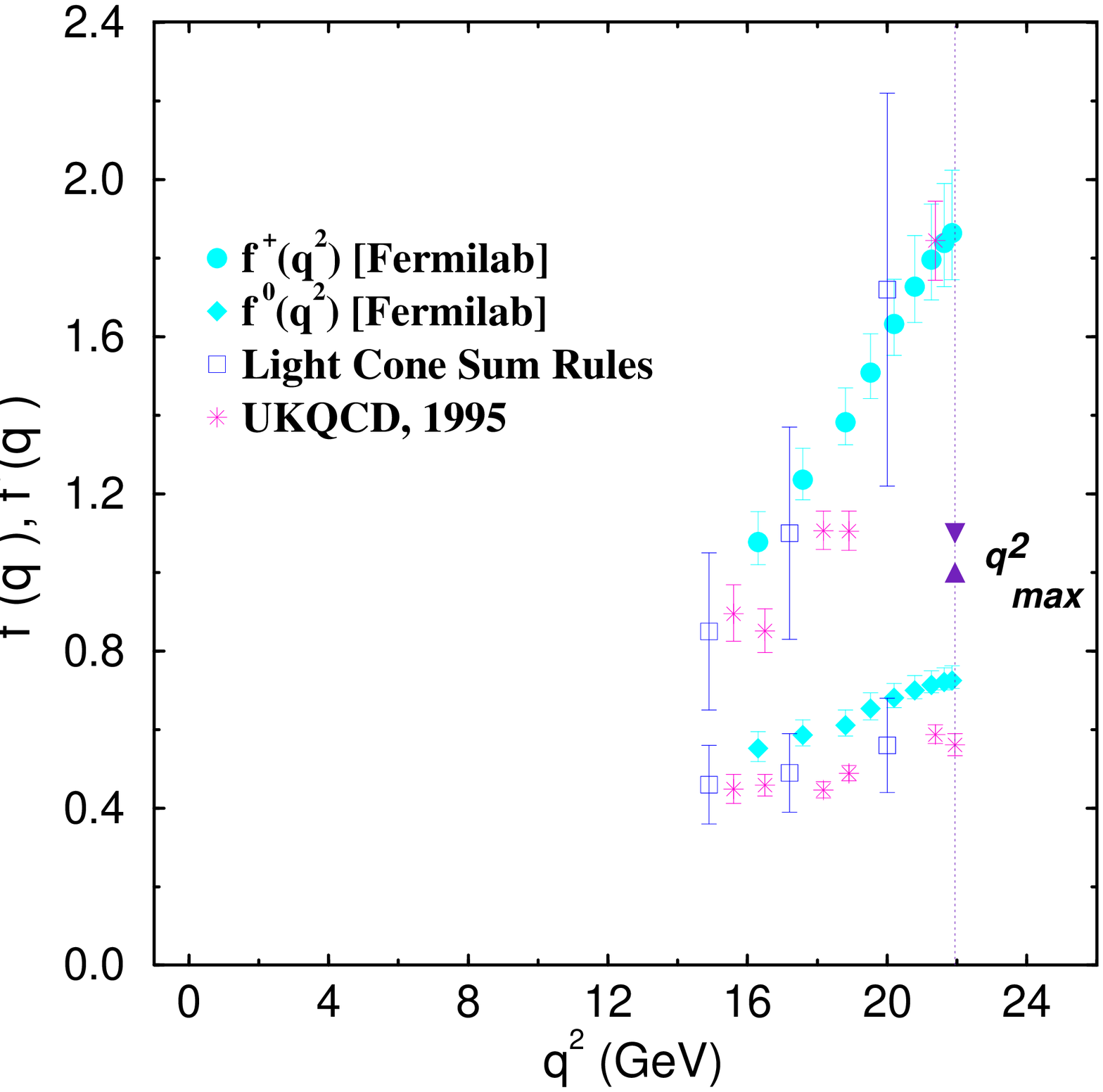}}\end{picture} }
\put(7.5,2){\begin{picture}(7,7)\put(-0.9,-0.4)
{\shit{6.5cm}{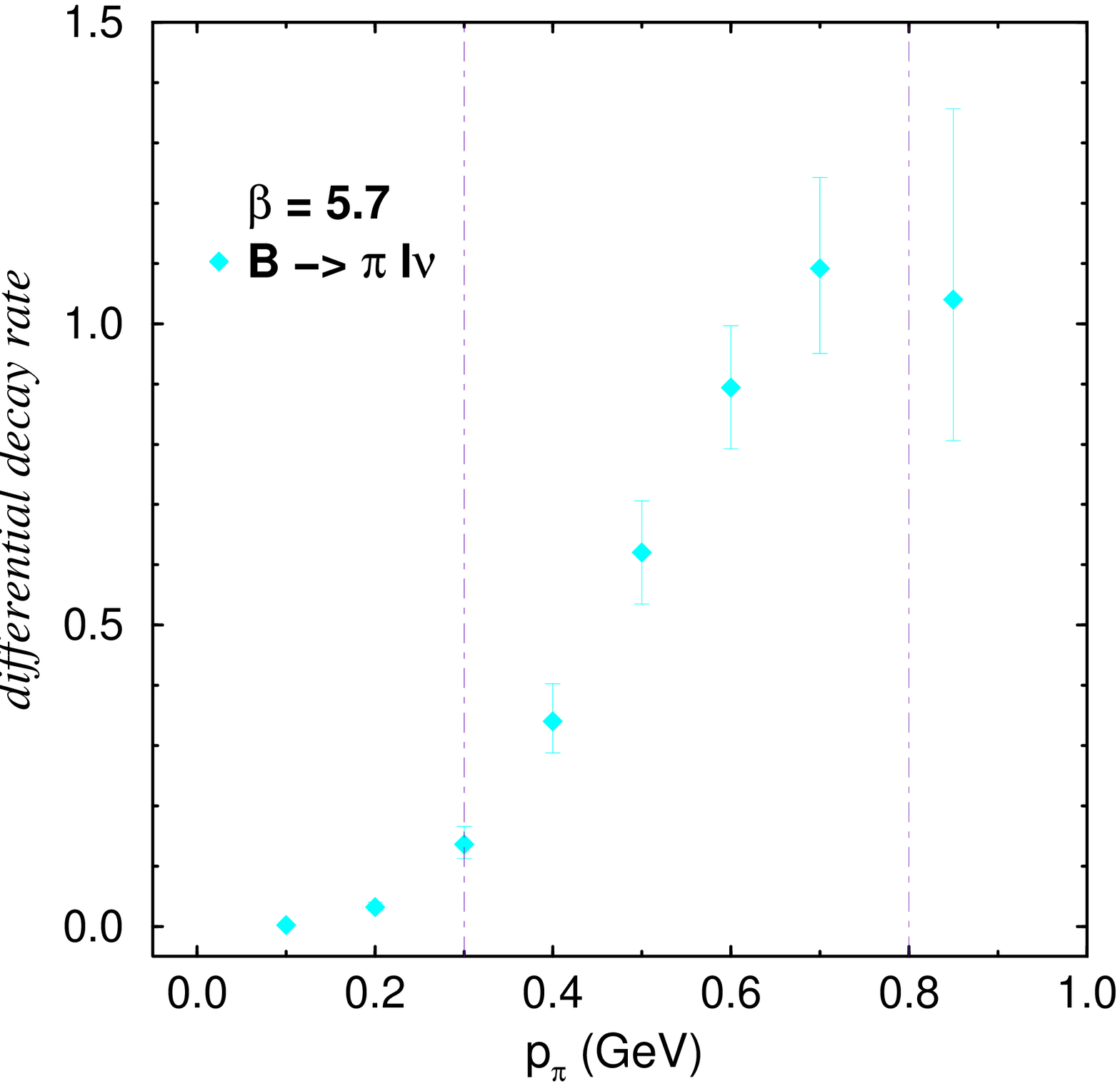}}\end{picture}}
\end{picture}
\end{center}
\vskip -.2 cm
\caption[]{
\label{ff_and_dG}
\small The $q^2$ dependence of the form factors by the Fermilab group for strange light quarks and $a\rightarrow 0$, compared with results from UKQCD~\cite{UKQCD95} and Light cone sum rules~\cite{pball}. The lattice data only include statistical errors. The right-hand plot is the decay rate (less the momentun independent prefactors) versus pion momentum. The dotted lines define a region in pion momentum where it is believed the lattice calculation is most reliable and for which there are experimental data. } 
\end{figure}

To determine $|V_{ub}|$ the required parameters are the form factors, $f^{+,0}(q^2)$ given by
\begin{equation}
\langle P^\prime (\vec k)|V_\mu |P(\vec p )\rangle = \left ( p+k-q\cdot\frac{m_P^2-m_{P^\prime}^2}{q^2}\right )_\mu f^+(q^2) + q_\mu\frac{m_P^2-m_{P^\prime}^2}{q^2}f^0(q^2) ,
\end{equation}
where $q^2$ is the momentum transfer ie. $q_\mu = (p-k)_\mu$.
These decays are trickier than the leptonic case because the final state hadron introduces an extra kinematic parameter - the momentum of the final state particle. Lattice calculations work best at small three-momenta because 
momentum-dependent errors are under control ie. close to $q^2_{ \mbox{max}} = (m_P-m_{P^\prime})^2$. However, it is traditional to quote
form factors at $q^2=0$ since at this point experiments can measure a slope and
intercept for the form factor, so a large model-dependent extrapolation in $q^2$ is performed. In contrast, for $D\rightarrow\pi l\nu$ the entire kinematic range can be covered, eg. Ref.~\cite{simone_dpf}. 
Figure~\ref{ff_and_dG} shows preliminary data from Fermilab for the form factors as a function of $q^2$. For the first time the lattice spacing dependence of these matrix elements has been studied and for $300\mbox{MeV}\leq p_\pi\leq 800\mbox{MeV}$ this is found to be mild~\cite{my_lat98}. 
Finally I note that $|V_{ub}|$ can be determined without recourse to a $q^2$-extrapolation by calculating partial widths from the differential decay rate 
\begin{equation}
\left .\frac{d\Gamma}{d|p_\pi |}\right |_{300MeV\leq p_\pi\leq 800MeV} = \frac{2m_BG_F^2|V_{ub}|^2}{24\pi^3}\frac{|p_\pi |^4}{E_\pi}|f^+(q^2)|^2 ,
\end{equation}
in a region where theory and experiment have reliable data, as in Figure~\ref{ff_and_dG}.
\subsection{$B\longrightarrow Dl\nu$}
The final topic is a new development in the calculation of form factors at zero recoil for the decay $B\rightarrow D^{(\ast )}l\nu$. The matrix element is 
parameterised by two form factors, $h^{(+,-)}(1)$ which are combined in a physical form factor, \\\centerline{${\cal F}_{B\rightarrow D}(1) = h_+^{B\rightarrow D}(1) -  h_-^{B\rightarrow D}(1) (m_B-m_D)/(m_B+m_D) $ .}\\
$|V_{cb}|$ is determined from experimental data for ${\cal F}(1)|V_{cb}|$.
Although the shape of the form factor has been calculated successfully (see eg. Refs.~\cite{flynn,draper}) lattice calculations of the absolute value at zero recoil were frought with difficulties both statistical and systematic. But it has recently been shown that by extracting the form factors from a {\it ratio} of matrix elements at zero recoil, the bulk of the uncertainties cancel leaving an extremely precise result for the form factor~\cite{shoji} and therefore $|V_{cb}|$. For $h^+(1)$ one constructs the ratio
\begin{equation}
\frac{\langle D|V_0|B\rangle\langle B|V_0|D\rangle}{\langle D|V_0|D\rangle\langle B|V_0|B\rangle} = \frac{h_+^{B\rightarrow D}(1)h_+^{D\rightarrow B}(1)}{h_+^{D\rightarrow D}(1)h_+^{B\rightarrow B}(1)} = |h_+^{B\rightarrow D}(1)|^2 .
\end{equation}
There is a similar expression using double ratios for $h^-(1)$. Preliminary results~\cite{shoji} show this is indeed very successful,
\begin{equation}
{\cal F}(1) = 1.069\pm 0.008\pm 0.002\pm 0.025 .
\end{equation}
In fact this method has the added advantage that at zero recoil it is a deviation of ${\cal F}(1)$ from 1 that is measured so that effects of quenching and lattice spacing dependence which do not cancel in the ratio are expected to be small.


\begin{references}  % All references should follow standard format
%
\bibitem{PDG}Particle Data Group, C. Caso \etal, The European Physical Journal C3 (1998) 1  
%
\bibitem{draper}T.~Draper, hep-lat/9810065, plenary talk given at Lattice'98, Boulder CO, USA. 
%
\bibitem{askHQ_proc}A.~Kronfeld, hep-ph/9812288, AIP Conference Proceedings {\bf 459} Heavy Quarks at Fixed Target, 355 (1998) and references therein.
%
\bibitem{JLQCDfB}S.~Aoki \etal (JLQCD Collaboration), hep-lat/9711041.
%
\bibitem{fB_paper}A.~El-Khadra, A.~Kronfeld, P.~Mackenzie, S.~Ryan, J.~Simone, Phys.~Rev.~{\bf D58} 014506 (1998).
%
\bibitem{GLOKfB} A.~Ali Khan {\it et al}, hep-lat/9801038, Phys.Lett. {\bf B427} 132-140 (1998).
%
\bibitem{MILCfB}C.~Bernard \etal (MILC Collaboration),  Phys.Rev.Lett. {\bf 81} 4812-4815 (1998).
%
\bibitem{UKQCDfBdgr}D.~Richards \etal (UKQCD Collaboration), hep-lat/9809064, to appear in the proceedings of Lattice'98.
%
\bibitem{UKQCDfBlin}L.~Lellouch and C.-J.~D.~Lin (UKQCD Collaboration), hep-lat/9809018, to appear in the proceedings of Lattice'98.
%
\bibitem{APEfB}D.~Becirevic {\it et al} (APE Collaboration), hep-lat/9811003.
%
\bibitem{NRQCD}W.E.~Caswell and G.P.~Lepage, Phys. Lett. {\bf B167} 437 (1986); G.P.~Lepage and B.A.~Thacker, Nucl. Phys. {\bf B} Proc. Suppl.{\bf 4} 199 (1987);  B.A.~Thacker and G.P.~Lepage, Phys. Rev. {\bf D43} 196 (1991); G.P.~Lepage, L.~Magnea, C.~Nakhleh, U.~Magnea and K.~Hornbostel, Phys. Rev. {\bf D46} 4052 (1992).
%
\bibitem{KKM}A.~El-Khadra, A.~Kronfeld and P.~Mackenzie, Phys. Rev. {\bf D55} 3933 (1997).
%
\bibitem{my_lat98}S.~Ryan \etal hep-lat/9810041, to appear in the proceedings of Lattice'98, Boulder CO, USA.
%
\bibitem{alpha_collab}M.~L\"uscher, S.~Sint, R.~Sommer and P.~Weisz, Nucl. Phys. {\bf B478} 365 (1996).
%
\bibitem{WA75}WA75 Collaboration (S. Aoki, \etal.), Prog. Th. Phys. 89, 131 (1993).
%
\bibitem{CLEO}CLEO (D. Acosta \etal.) Phys. Rev. D49, 5690 (1994); CLEO (D. Bliss \etal ) Phys.Rev. {\bf D57} 5903 (1998). 
%
\bibitem{E653}E653 Collaboration (K. Kodama, \etal.), Phys. Lett. B382, 299 (1996).
%
\bibitem{L3}L3 Collaboration (M. Acciarri, \etal.), Phys. Lett. B396, 327 (1997).
%
\bibitem{DELPHI}DELPHI Collaboration (F. Parodi, \etal.), International Europhysics Conference on High Energy Physics, Jerusalem 19-26 Aug 1997.
%
\bibitem{aoki_pertTh}K-I.~Ishikawa, S.~Aoki, S.~Hashimoto, H.~Matsufuru, T.~Onogi, N.~Yamada, Nucl.Phys.Proc.Suppl. 63 344 (1998).
%
\bibitem{bernard_nf2}C.~Bernard, private communication.
%
\bibitem{sara_nf2}S.~Collins {\it et al}, hep-lat/9901001.
%
\bibitem{simone_dpf}J.~Simone, these proceedings.
%
\bibitem{UKQCD95}UKQCD Collaboration, Nucl. Phys. {\bf B447} 425 (1995).
%
\bibitem{pball}P.~Ball, hep-ph/9802394, JHEP 9809, 005(1998) 
%
\bibitem{flynn}J.~Flynn and C.~Sachrajda, hep-lat/9710057, Heavy Flavours (2nd edition) edited by A.~Buras and M.~Lindner (World Scientific, Singapore).
\bibitem{shoji}S.~Hashimoto \etal hep-lat/9810056, to appear in the proceedings of Lattice'98, Boulder CO, USA.
%
\end{references}
\end{document}